\documentclass[]{spie}  
\usepackage{graphicx}
\usepackage{rotating}
\usepackage{amssymb}
\usepackage{mathptmx}
\usepackage{epstopdf}
\usepackage[]{natbib}
\usepackage{lineno}
\usepackage{url}

\bibpunct{[}{]}{,}{n}{}{,}

\begin{document}

\newcommand{\hdblarrow}{H\makebox[0.9ex][l]{$\downdownarrows$}-}
\newcommand{\pb}{\protect\textsc{polarbear}}
\newcommand{\Pb}{\protect\textsc{Polarbear}}

\title{The LiteBIRD Satellite Mission - Sub-Kelvin Instrument}

\author{  
A. Suzuki $^{44*}$, 
P. A. R. Ade $^{48}$, 
Y. Akiba $^{19,50}$,  		
D. Alonso $^{42}$,  
K. Arnold $^{16}$, 	
J. Aumont $^{20}$, 
C. Baccigalupi $^{25}$, 
D. Barron $^{49}$, 
S. Basak $^{11,25}$,  
S. Beckman $^{15}$
% ...
\skiplinehalf
\small{
% ...
$^{11}$Indian Institute of Science Education and Research, Vithura, Thiruvananthapuram - 695551, India\\
% ...
$^{15}$Department of Physics, University of California, Berkeley, CA 94720, USA\\ 
$^{16}$Department of Physics, University of California, San Diego, CA 92093-0424, USA\\ 
% ...
$^{19}$High Energy Accelerator Research Organization (KEK), Tsukuba, Ibaraki 305-0801, Japan\\ 
$^{20}$Institut d'Astrophysique Spatiale (IAS), CNRS, UMR 8617, Universit$\acute{\rm e}$ Paris-Sud 11, B$\hat{\rm a}$timent 121, 91405 Orsay, France\\ 
% ...
$^{25}$International School for Advanced Studies (SISSA), Via Bonomea 265, 34136, Trieste, Italy\\ 
% ...
$^{42}$Oxford Astrophysics, Oxford, OX1 3RH, United Kingdom\\ 
% ...
$^{44}$Physics Division, Lawrence Berkeley National Laboratory, Berkeley, CA 94720, USA\\ 
% ...
$^{48}$School of Physics and Astronomy, Cardiff University, Cardiff CF10 3XQ, United Kingdom\\ 
$^{49}$Space Sciences Laboratory, University of California, Berkeley, CA 94720, USA\\ 
$^{50}$The Graduate University for Advanced Studies (SOKENDAI), Miura District, Kanagawa 240-0115, Hayama, Japan\\ 
% ...
$^{*}$ Corresponding author: asuzuki@lbl.gov 
}
}


\maketitle

\begin{abstract}
Inflation is the leading theory of the first instant of the universe. 
Inflation, which postulates that the universe underwent a period of rapid expansion an instant after its birth, provides convincing explanation for cosmological observations.
Recent advancements in detector technology have opened opportunities to explore primordial gravitational waves generated by the inflation through ``B-mode'' (divergent-free) polarization pattern embedded in the Cosmic Microwave Background anisotropies. 
If detected, these signals would provide strong evidence for inflation, point to the correct model for inflation, and open a window to physics at ultra-high energies. 

LiteBIRD is a satellite mission with a goal of detecting degree-and-larger-angular-scale B-mode polarization.
LiteBIRD will observe at the second Lagrange point with a 400 mm diameter telescope and 2,622 detectors.
It will survey the entire sky with 15 frequency bands from 40 to 400 GHz to measure and subtract foregrounds.

The U.S. LiteBIRD team is proposing to deliver sub-Kelvin instruments that include detectors and readout electronics.
A lenslet-coupled sinuous antenna array will cover low-frequency bands (40 GHz to 235 GHz) with four frequency arrangements of trichroic pixels.
An orthomode-transducer-coupled corrugated horn array will cover high-frequency bands (280 GHz to 402 GHz) with three types of single frequency detectors.
The detectors will be made with Transition Edge Sensor (TES) bolometers cooled to a 100 milli-Kelvin base temperature by an adiabatic demagnetization refrigerator.
The TES bolometers will be read out using digital frequency multiplexing with Superconducting QUantum Interference Device (SQUID) amplifiers. 
Up to 78 bolometers will be multiplexed with a single SQUID ampliﬁer. 

We report on the sub-Kelvin instrument design and ongoing developments for the LiteBIRD mission.

\end{abstract}

\section{Introduction}
Over past decades scientists built millimeter-wave sensitive instruments to characterize the Cosmic Microwave Background (CMB) from the ground, balloon, and satellite. 
Measurement of the uniformity of the CMB temperature is a pillar of the big bang theory, and exquisite measurement of its anisotropy is beautifully modeled by the $\Lambda$CDM model with just six free parameters. 
CMB measurements unveiled that the universe is geometrically flat, has Dark Energy, and other fascinating facts. 
Intriguing questions emerged from these findings.
Why is the universe homogeneous over such large distance? 
Why is the universe geometrically flat?
What seeded fluctuations that started galaxy clusters?
Inflation is the leading theory that provides answers to these questions.
It postulates that the universe underwent a period of rapid expansion an instant after its birth.
The theory predicts that rapid expansion during the period generated gravitational waves, which would then imprint polarization on the CMB through Thomson scattering. 
Scalar perturbations that are responsible for temperature anisotropy and structure formation could only produce the parity-conserving polarization pattern (E-mode). 
Graviational waves would imprint both parity-conserving and parity-violating (B-mode) polarization patterns in the CMB at degree angular scales.
Deflection of the CMB by weak gravitational lensing also produces B-mode by modifying E-mode patterns. 
However, this lensing B-mode appears at arcmin angular scales.
This lack of B-mode polarization from primordial scalar perturbations and ability to separate lensing B-mode through angular scale make large angular size B-mode polarization an ideal signature to seek from the inflationary epoch. 
However, the primordial B-mode signal from the inflation is expected to be faint. 
The CMB temperature anisotropy is about 80 $\mu\mathrm{K}$, whereas the polarization from the inflationary gravitational waves is expected to be less than 100nK.
In addition, galactic foregrounds, synchroton radiation, and thermal dust emission emit polarized foreground signals to make the search for primordial B-modes more challenging.
Recent advancements in detection technology have opened a path to overcome these challenges in the quest for the primordial B-mode.

\section{Mission Overview}
LiteBIRD is a next generation CMB polarization satellite designed to probe the inflationary B-modes.
The science goal of LiteBIRD is to measure the tensor-to-scalar ratio ($r$) with a sensitivity of $\delta r < 0.001$  to explore the major large-single-field slow-roll inflation models. 
LiteBIRD will target B-mode polarization on angular scales of a degree and more, where that primordial B-mode signature is expected to peak.
JAXA's H3 launch vehicle will deploy the LiteBIRD satellite to the second Lagrange point.
LiteBIRD will observe for three years. 
LiteBIRD will survey the entire sky with 2,622 detectors using a 
spin and precession strategy.
Instrument is designed to achieve 4.1 $\mu\mathrm{K}\cdot\mathrm{arcmin}$ noise equivalent CMB temperature at 140 GHz.
The LiteBIRD will cover 15 frequency bands from 40 to 400 GHz to measure and subtract galactic foregrounds.
Sensitivities for other frequency bands are shown in Figure~\ref{fig:coldsystem}.

\section{The Instrument Overview}
\begin{figure}[!h]
\begin{center}
\includegraphics[width=\textwidth,keepaspectratio]{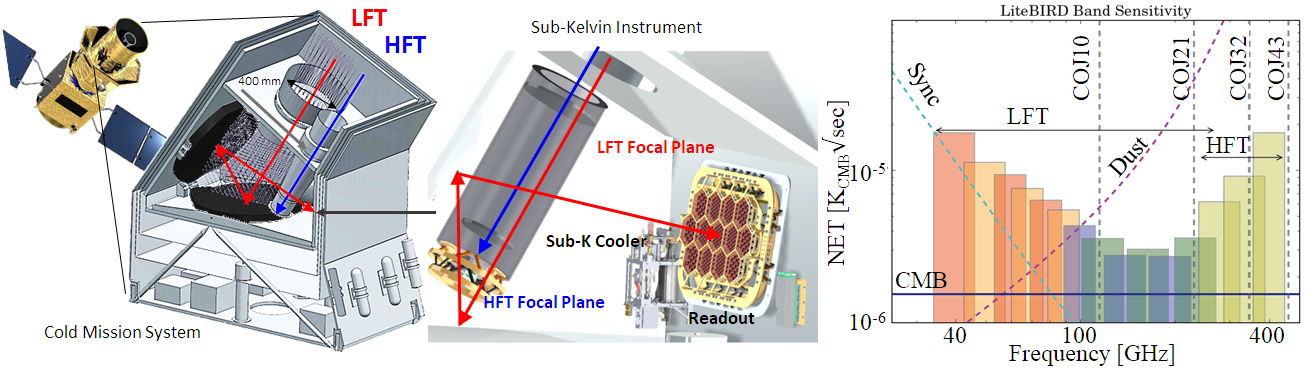}
\end{center}
\caption{(Color online) (Left) Cross section of the cold mission system and insert, highlighting the sub-Kelvin instrument. The red and blue arrows indicate the optical path for the LFT and the HFT, respectively. (Right) The LiteBIRD sensitivities for each frequency band. Different colors indicate different focal plane modules. The LFT has four types of modules with a staggering frequency schedule. The HFT has three single color pixel types with overlapping frequency coverage.}
\label{fig:coldsystem}
\end{figure}

LiteBIRD satellite is organized into two systems. 
The mission BUS system handles telemetry and power management tasks.
The cold mission system houses scientific instruments.
Inside the cold mission system, there are two telescopes, as shown in Figure~\ref{fig:coldsystem}.
LiteBIRD covers 40 GHz - 400 GHz using two telescopes. 
The Low Frequency Telescope (LFT) covers 40 GHz to 235 GHz, and the High Frequency Telescope (HFT) covers 280 GHz to 400 GHz.
LFT has a 400 mm aperture Crossed-Dragone telescope, and HFT has a 200 mm aperture on-axis refractor with two silicon lenses.
Both telescopes will have a cryogenic rotating achromatic half-wave plate at their apertures. 
Half-wave plates are used to modulate the polarization signal to suppress 1/f noise and mitigate systematic bias associated with beam and polarization non-ideality \cite{TomoLTD17,ABSHWP}.
The combination of Stirling and Joule Thomson coolers provide cryogenic stages above 2 Kelvin.
The sub-Kelvin instrument section consists of detectors, readout electronics, and a sub-kelvin cooler.
The focal plane is cooled to 100 mK with a two stage sub-Kelvin cooler.
The TES bolometer detector arrays are read out with digital frequency multiplexing.

\section{Sub-Kelvin Instrument}
\begin{figure}[!h]
\begin{center}
\includegraphics[width=\textwidth,keepaspectratio]{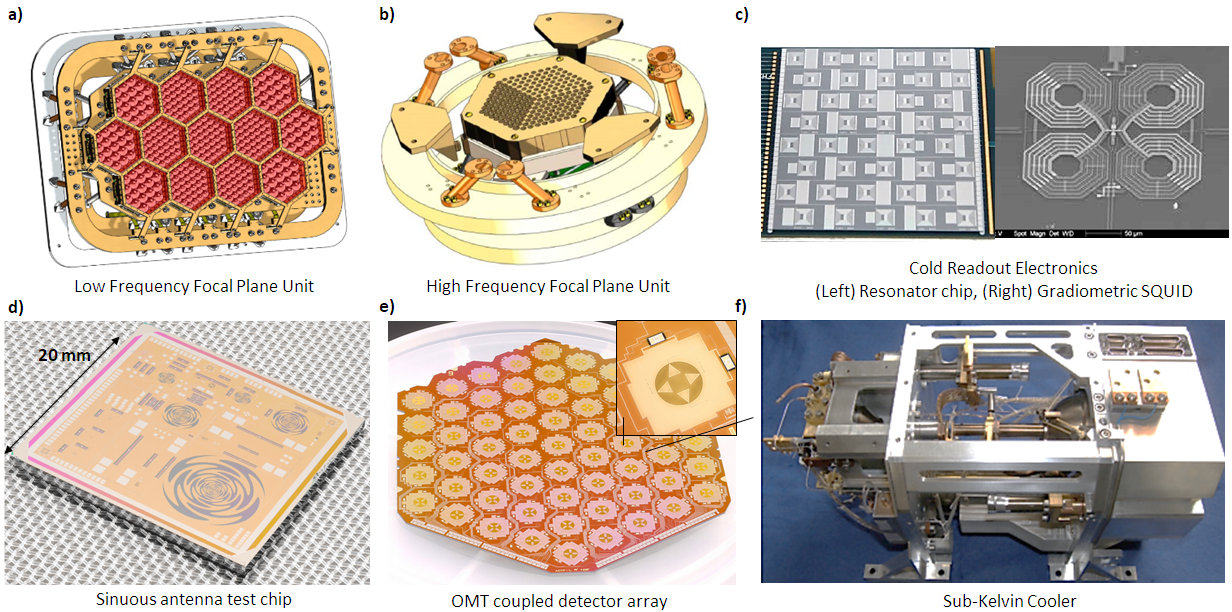}
\end{center}
\caption{(Color online) a) CAD drawing of the LFU. The unit is 420 mm x 600 mm. The design is based on the POLARBEAR focal plane. b) CAD drawing of the HFU. The unit is 260 mm diameter. The design is based on the ACTpol focal plane. c) Photograph of a superconducting resonator chip (left) and a gradiometric SQUID (right) for the cryogenic readout electronics. d) Photograph of a test chip fabricated for LiteBIRD detector demonstration. The multiple sinuous antenna coupled detector covers 30 GHz to 360 GHz. e) Photograph of an OMT coupled detector array fabricated at NIST. f) Photograph of CEA's two stage sub-Kelvin cooler. The unit is 300 mm across.
\vspace{-4 mm}}
\label{fig:subkelvin}
\end{figure}

The Sub-Kelvin instrument is organized into four systems: the low frequency focal plane unit (LFU), the high frequency focal plane unit (HFU), cryogenic readout electronics (CRE), and the sub-Kelvin cooler. 
The design and representative parts are shown in Figure~\ref{fig:subkelvin}.
The U.S. LiteBIRD team will deliver the LFU, HFU, and CRE.
The sub-Kelvin cooler will be provided by CEA.

The LFU covers frequency bands from 40 GHz to 235 GHz.
The detector technology for the LFU is multi-chroic lenslet coupled sinuous antenna detectors that have been adapted by multiple sub-orbital experiments (Simons Array, SPT-3G, Simons Observatory) \cite{SuzukiLTD15, WestbrookLTD17, SPT3GLTD17}.
Multichroic detectors are implemented to cover the wide frequency range of the LFU while keeping focal plane size compact. 
The LFU is further split into four detector modules that have different pixel sizes and frequency coverage. 
Low frequency modules have 18 mm diameter pixels that cover 40 GHz to 90 GHz with two frequency arrangements of triplexed filters.
Mid frequency modules have 12 mm diameter pixels that cover 100 GHz to 235 GHz with two frequency arrangements of triplexed filters.
Frequency coverages of the modules are slightly shifted from each other to cover 40 GHz to 235 GHz with 12 distinct bands as shown in Figure~\ref{fig:coldsystem}. 
The detector module and focal plane design is based on the POLARBEAR experiment.

The HFU covers 280 GHz, 337 GHz, and 402 GHz channels.
The detector technology for the HFU is single color an orthomode-transducer (OMT) coupled corrugated horn detectors that were deployed on multiple sub-orbital experiments such as ACT-pol and SPT-pol, and will be deployed on SPIDER \cite{yoon2009a, Austermann2012, hubmayr2016}.
OMT-coupled corrugated horn technology was chosen for HFU for its high technology readiness level at high frequency.
Corrugated horn was chosen over spline-profiled horn for symmetry of beam and cleanliness of its polarization property.
Corrugated horns will be fabricated from gold plated silicon platelette arrays.
The detector module and focal plane design is based on the ACTpol and SPTpol experiments.

Fabrication of detectors for the LiteBIRD will be done at the NIST Boulder Microfabrication facility and the Marvell Nanofabrication laboratory at the University of California, Berkeley.
The facility at NIST is an 18,000 square foot class 100 clean room.  It has produced CMB detectors for multiple CMB experiments such as Advanced ACTpol, ACT-pol, SPT-pol, SPIDER and BLAST.
The facility at Berkeley is an 15,000 square foot class 100 clean room.  It also has produced CMB detectors for multiple CMB experiments, such as APEX-SZ, POLARBEAR-1, SPT-SZ, EBEX, POLARBEAR-2/Simons Array.
The TES bolometers will be fabricated with aluminum-manganese alloy with superconducting critical temperature tuned to 170 milli-Kelvin \cite{Dale}.

The TES bolometers will be read out with digital frequency multiplexing \cite{Haan2012}.
This technology has been adapted by multiple sub-orbital experiments such as Simons Array and SPT-3G.
An FPGA electronics board at a warmer stage will generate bias currents with specific frequency tones.
The FPGA board that will be used by LiteBIRD is specifically developed for radiation hardness by McGill university in Canada.
Frequency multiplexing will be done with micro-fabricated superconducting resonators.
First stage amplification will be done by Gradiometric series SQUID arrays at the 2 Kelvin stage.
Up to 78 TES bolometers will be read out by a single SQUID amplifier to reduce the heat load on the milli-Kelvin stage and to reduce power consumption and parts count on warmer stages.

Although instruments for sub-orbital experiments and satellite missions are similar, there are many details that are unique a satellite mission.
The LiteBIRD sub-Kelvin instrument team has been testing components specifically for a space environment. 
The instrument in a satellite environment can achieve high sensitivity due to the satellite's low background optical power environment.
For a CMB satellite experiment, some detectors will receive an order of magnitude lower optical power than ground based experiments.
We successfully fabricated and tested aluminum-manganese TES bolometers with a critical temperature of 180 milli-Kelvin to achieve sub-pW saturation power at 100 mK base temperature.
The Planck satellite reported that glitches in the data stream due to cosmic rays had significant impact on its data analysis \cite{2014A&A...569A..88C}.
The US LiteBIRD team is collaborating with LiteBIRD collaborators in Japan to study cosmic ray interaction with detector parts using GEANT-4 simulations.
We are also testing the effect of various glitch mitigation features on detector wafers with a radioactive source inside a cryostat.
To mitigate glitch amplitude, we are engineering a high impedance phonon barrier to block propagation of phonons generated by cosmic rays to the TES bolometer. 
We are studying possible long term degradation of components from energetic cosmic rays by irradiating detector parts with 5 kilo-rad to 10 kilo-rad of energetic protons at the HIMAC facility in Japan.
We also conducted a launch survival test of the detector module by putting an assembled detector module through a shake test at the Space Science Laboratory.
The detector module survived 15 g of vibration during the test, and in addition there was no measured resonance below 1 kHz.

The Sub-Kelvin cooler will be provided by CEA.
The CEA team had built sub-Kelvin coolers for the SPICA satellite mission \cite{1757-899X-101-1-012010}.
LiteBIRD's sub-Kelvin cooler has two temperature stages.
A He-3 adsorption fridge provides the 300 mK stage, and an Adiabatic Demagnetization Refrigerator with a CPA salt pill provides 100 mK stage.
The fridge is capable of providing a 25 hour hold time with the expected heat load from the LiteBIRD focal plane.
The cryogenic chain of LiteBIRD can achieve an 89\% duty cycle to maximize observing time.
The unit is designed with magnetic shielding to minimize the effect of magnetic fields on instrument components such as the SQUIDs and TES bolometers. 
The sub-Kelvin cooler is tested to withstand 21 g of vibration and a static acceleration of 120g.

\section{Conclusion}
LiteBIRD is a next generation CMB satellite to probe inflation with an uncertainty on the tensor to scalar ratio $\delta r<0.001$.
The LiteBIRD team in Japan is at Phase A1. The concept development study is in progress.
The U.S. LiteBIRD team is in a technology development phase to raise the Technology Readiness Level of sub-Kelvin instrument components.

\begin{acknowledgements}
Development for the sub-Kelvin instrument was supported by the NASA Mission of Opportunity Phase A effort.
\end{acknowledgements}



\begin{thebibliography}{10}

\bibitem{TomoLTD17}
T.~Matsumura et~al.
\newblock Development of a half-wave plate based polarization modulator unit
  for litebird.
\newblock {\em Journal of Low Temperature Physics This Special Issue}, 2017
  Submitted.

\bibitem{ABSHWP}
A.~Kusaka, T.~Essinger-Hileman, J.~W. Appel, P.~Gallardo, K.~D. Irwin,
  N.~Jarosik, M.~R. Nolta, L.~A. Page, L.~P. Parker, S.~Raghunathan, J.~L.
  Sievers, S.~M. Simon, S.~T. Staggs, and K.~Visnjic.
\newblock Modulation of cosmic microwave background polarization with a warm
  rapidly rotating half-wave plate on the atacama b-mode search instrument.
\newblock {\em Review of Scientific Instruments}, 85(2):024501, 2014.

\bibitem{SuzukiLTD15}
A.~Suzuki, K.~Arnold, J.~Edwards, G.~Engargiola, W.~Holzapfel, B.~Keating, A.T.
  Lee, X.F. Meng, M.J. Myers, R.~O’Brient, E.~Quealy, G.~Rebeiz, P.L.
  Richards, D.~Rosen, and P.~Siritanasak.
\newblock Multi-chroic dual-polarization bolometric detectors for studies of
  the cosmic microwave background.
\newblock {\em Journal of Low Temperature Physics}, 176(5-6):650--656, 2014.

\bibitem{WestbrookLTD17}
B.~Westbrook et~al.
\newblock The polarbear-2 and simons array focal plane fabrication status.
\newblock {\em Journal of Low Temperature Physics This Special Issue}, 2017
  Submitted.

\bibitem{SPT3GLTD17}
A.~Anderson et~al.
\newblock Spt-3g: A multichroic receiver for the south pole telescope.
\newblock {\em Journal of Low Temperature Physics This Special Issue}, 2017
  Submitted.

\bibitem{yoon2009a}
K.~W. {Yoon}, J.~W. {Appel}, J.~E. {Austermann}, J.~A. {Beall}, D.~{Becker},
  B.~A. {Benson}, L.~E. {Bleem}, J.~{Britton}, C.~L. {Chang}, J.~E.
  {Carlstrom}, H.-M. {Cho}, A.~T. {Crites}, T.~{Essinger-Hileman},
  W.~{Everett}, N.~W. {Halverson}, J.~W. {Henning}, G.~C. {Hilton}, K.~D.
  {Irwin}, J.~{McMahon}, J.~{Mehl}, S.~S. {Meyer}, S.~{Moseley}, M.~D.
  {Niemack}, L.~P. {Parker}, S.~M. {Simon}, S.~T. {Staggs}, K.~{U-Yen},
  C.~{Visnjic}, E.~{Wollack}, and Y.~{Zhao}.
\newblock {Feedhorn-Coupled TES Polarimeters for Next-Generation CMB
  Instruments}.
\newblock 1185:515--518, December 2009.

\bibitem{Austermann2012}
J.~E. Austermann, K.~A. Aird, J.~A. Beall, D.~Becker, A.~Bender, B.~A. Benson,
  L.~E. Bleem, J.~Britton, J.~E. Carlstrom, C.~L. Chang, H.~C. Chiang, H.-M.
  Cho, T.~M. Crawford, A.~T. Crites, A.~Datesman, T.~de~Haan, M.~A. Dobbs,
  E.~M. George, N.~W. Halverson, N.~Harrington, J.~W. Henning, G.~C. Hilton,
  G.~P. Holder, W.~L. Holzapfel, S.~Hoover, N.~Huang, J.~Hubmayr, K.~D. Irwin,
  R.~Keisler, J.~Kennedy, L.~Knox, A.~T. Lee, E.~Leitch, D.~Li, M.~Lueker,
  D.~P. Marrone, J.~J. McMahon, J.~Mehl, S.~S. Meyer, T.~E. Montroy, T.~Natoli,
  J.~P. Nibarger, M.~D. Niemack, V.~Novosad, S.~Padin, C.~Pryke, C.~L.
  Reichardt, J.~E. Ruhl, B.~R. Saliwanchik, J.~T. Sayre, K.~K. Schaffer,
  E.~Shirokoff, A.~A. Stark, K.~Story, K.~Vanderlinde, J.~D. Vieira, G.~Wang,
  R.~Williamson, V.~Yefremenko, K.~W. Yoon, and O.~Zahn.
\newblock Sptpol: an instrument for cmb polarization measurements with the
  south pole telescope.
\newblock {\em Proc.SPIE}, 8452:8452 -- 8452 -- 18, 2012.

\bibitem{hubmayr2016}
Johannes Hubmayr, Jason~E. Austermann, James~A. Beall, Daniel~T. Becker,
  Steven~J. Benton, A.~Stevie Bergman, J.~Richard Bond, Sean Bryan, Shannon~M.
  Duff, Adri~J. Duivenvoorden, H.~K. Eriksen, Jeffrey~P. Filippini, A.~Fraisse,
  Mathew Galloway, Anne~E. Gambrel, K.~Ganga, Arpi~L. Grigorian, Riccardo
  Gualtieri, Jon~E. Gudmundsson, John~W. Hartley, M.~Halpern, Gene~C. Hilton,
  William~C. Jones, Jeffrey~J. McMahon, Lorenzo Moncelsi, Johanna~M. Nagy,
  C.~B. Netterfield, Benjamin Osherson, Ivan Padilla, Alexandra~S. Rahlin,
  B.~Racine, John Ruhl, T.~M. Rudd, J.~A. Shariff, J.~D. Soler, Xue Song,
  Joel~N. Ullom, Jeff~Van Lanen, Michael~R. Vissers, I.~K. Wehus, Shyang Wen,
  D.~V. Wiebe, and Edward Young.
\newblock Design of 280 ghz feedhorn-coupled tes arrays for the balloon-borne
  polarimeter spider.
\newblock {\em Proc.SPIE}, 9914:9914 -- 9914 -- 14, 2016.

\bibitem{Dale}
Dale Li, Jason~E. Austermann, James~A. Beall, Daniel~T. Becker, Shannon~M.
  Duff, Patricio~A. Gallardo, Shawn~W. Henderson, Gene~C. Hilton, Shuay-Pwu Ho,
  Johannes Hubmayr, Brian~J. Koopman, Jeffrey~J. McMahon, Federico Nati,
  Michael~D. Niemack, Christine~G. Pappas, Maria Salatino, Benjamin~L. Schmitt,
  Sara~M. Simon, Suzanne~T. Staggs, Jeff Van~Lanen, Jonathan~T. Ward, and
  Edward~J. Wollack.
\newblock Almn transition edge sensors for advanced actpol.
\newblock {\em Journal of Low Temperature Physics}, 184(1):66--73, Jul 2016.

\bibitem{Haan2012}
Tijmen de~Haan, Graeme Smecher, and Matt Dobbs.
\newblock Improved performance of tes bolometers using digital feedback.
\newblock {\em Proc.SPIE}, 8452:8452 -- 8452 -- 10, 2012.

\bibitem{2014A&A...569A..88C}
A.~{Catalano}, P.~{Ade}, Y.~{Atik}, A.~{Benoit}, E.~{Br{\'e}ele}, J.~J. {Bock},
  P.~{Camus}, M.~{Chabot}, M.~{Charra}, B.~P. {Crill}, N.~{Coron},
  A.~{Coulais}, F.-X. {D{\'e}sert}, L.~{Fauvet}, Y.~{Giraud-H{\'e}raud},
  O.~{Guillaudin}, W.~{Holmes}, W.~C. {Jones}, J.-M. {Lamarre},
  J.~{Mac{\'{\i}}as-P{\'e}rez}, M.~{Martinez}, A.~{Miniussi}, A.~{Monfardini},
  F.~{Pajot}, G.~{Patanchon}, A.~{Pelissier}, M.~{Piat}, J.-L. {Puget},
  C.~{Renault}, C.~{Rosset}, D.~{Santos}, A.~{Sauv{\'e}}, L.~D. {Spencer}, and
  R.~{Sudiwala}.
\newblock {Impact of particles on the Planck HFI detectors: Ground-based
  measurements and physical interpretation}.
\newblock {\em AAP}, 569:A88, September 2014.

\bibitem{1757-899X-101-1-012010}
J-M Duval, L~Duband, and A~Attard.
\newblock Qualification campaign of the 50 mk hybrid sorption-adr cooler for
  spica/safari.
\newblock {\em IOP Conference Series: Materials Science and Engineering},
  101(1):012010, 2015.

\end{thebibliography}
\end{document}